\documentclass[letterpaper]{article}
\usepackage{verbatim}
\usepackage{graphicx}
\usepackage{amssymb}
\usepackage{amsmath}
\usepackage{newalg}
\usepackage{algorithmic}
\usepackage{algorithm,enumerate}
\usepackage{amsfonts}

\begin {document}
\title{Ownership Authentication Transfer Protocol for Ubiquitous Computing Devices}
\author{Pradeep B.H and Sanjay Singh\thanks {Sanjay Singh is with the Department of Information and Communication Technology, Manipal Institute of Technology, Manipal University, Manipal-576104, INDIA, E-mail: sanjay.singh@manipal.edu}}

\maketitle
\begin{abstract}
In ubiquitous computing devices, users tend to store some valuable information in their device. Even though the device can be borrowed by the other user temporarily, it is not safe for any user to borrow or lend the device as it may result the private data of the user to be public. To safeguard the user data and also to preserve user privacy we propose the technique of ownership authentication transfer. The user who is willing to sell the device has to transfer the ownership of the device under sale. Once the device is sold and the ownership has been transferred, the old owner will not be able to use that device at any cost. Either of the users will not be able to use the device if the process of ownership has not been carried out properly. This also takes care of the scenario when the device has been stolen or lost, avoiding the impersonation attack. The proposed protocol has been modeled and verified using Automated Validation of Internet Security Protocols and Applications (AVISPA) and is found to be safe.
\end{abstract}

\section{Introduction}
A ubiquitous computing (Ubicomp) or pervasive computing environment is imagined as a system with numerous invisible computers, sensors and actuators interacting with the user devices such as PDAs, Laptops, Mobile Phones etc. Data about the individuals who are a part of the ubiquitous environment is constantly being generated, transmitted, manipulated and stored. The user data present in the environment (in device or servers) is very sensitive. Protecting private data of every user in the environment is a major concern. Also in the this era of the mobile environment the user owns more than one portable devices like the PDAs, Laptops, Mobile Phones etc. with varying computing capabilities in order to access the variety of services that are being provided by the service providers. At times the user may tend to sell the device he/she owns. Since the device consists of the valuable information of the user and also will have the access to the valuable information present at the server, care should be taken to delete or remove the information of the previous owner and store the details of the new owner in the device as well as in the server. In spite of our best efforts, we were not able to find any similar protocols other than those cited below as part of literature review.  
\par
Paulo Tam and Jan Newmarch \cite{ptjn} in their work have suggested the concept of transferring the ownership of the device. The owner (old owner) of the device will send the message to the device itself that it is being bought by the other user (new owner). The device will send the message to the new owner saying that its ownership is about to change to you (new user), do you accept or reject. The new owner sends the response to the device. And the object will in turn send an acknowledgment on the status of the transfer to the old owner. However when the owner of the device is selling the device to the new owner, sending the message to the device itself does not seem feasible. Moreover to which device of the user, the device under sale is sending the message is not known. It is agreeable if the new owner of the device has one more device under his ownership. But if the user has no other device previously and it is his/her first device then there is no possibility for the device under sale to send the message to its new owner asking his consent on the ownership transfer. In ubiquitous environment the ownership transfer has to be informed to the central server instead of informing to the device under sale. 
\par
Jurgen Bohn \cite{jb} has mentioned that the user can borrow or lend the device to his friend or the stranger. The data of a particular user can be retrieved from the instant personalization server at any time and from anywhere for a specific time. Once the time limit is exceeded, the session will be ended and the user needs to quit the session or restart it. After using the device, the user can release the device and return it back to the owner of the device. But the very basic idea of sharing the personal device with a friend or a stranger may cause information to be public. This could be due to the other user being malicious (intentionally causing harm) by installing some kind of software which can record the data of the user or simply careless (unintentionally installing malicious software which can save the users data). Also due attention should be paid to the fact that the device could come with old data, if the transfer is incomplete due to technical reasons such as network congestion or lack of connectivity. The owner of the device may also turn out to be malicious with respect to the other user. The user may install a software that records the all the data that has been retrieved and sent from that device before encryption and after decryption. Later the user may be subjected to the impersonation attack. Moreover when the time limit is exceeded, there are chances that the user may have to end the session or restart it due to network latencies or unresponsive server when the user is trying to retrieve or release the data. 
\par
Yongming Jin et al \cite{yhz} has described the transfer of RFID from the old owner to the new owner. They define a protocol to safeguard the privacy of the respective owners by putting the clean stop between the transactions of the old and the new owners by means of a secret. The authors have suggested the use of RFIDs for the ownership transfer. But there are many security concerns with respect to the RFID tags. One of the primary RFID security concern is the illicit tracking of RFID tags. The tags are read by anyone in the world. If the person who read the tag is malicious can pose a risk by either impersonating the user or trying to manipulate the user data and use it for a wrong purpose. RFIDs working at a shorter range are vulnerable to skimming and eavesdropping. Even though certain RFID tags use cryptographic features, the cost and power requirements are very high when compare to the simpler RFID tags. Thus, the cost and power limitation has compelled some manufacturers to implement cryptographic tags using substantially weak encryption schemes, which are weak to resist the sophisticated attack. Moreover, the power available in the handheld devices is limited; these tags cannot be incorporated in the devices.
\par
Abdullah M. Alaraj \cite{aa} in his paper says that the users have to go to some official office for buying or selling the merchandise and to complete the process of ownership transfer. He also makes an assumption that the certain equipments are required for ownership transfer and tries to improve the fairness by including the transfer of money through the bank servers. But going to the an official office that deals with buying or selling of merchandise is suitable only to the goods like cars or for real estate. This scenario will not be suitable when applied to ubiquitous computing devices. The process of ownership transfer requires only a central key server and a device meant for sale. Submitting your bank details to the third party might be risky at the time of payment. Even thought if the system provides the best servers for transaction and promotes the users to submit their bank details to the device in an office meant for buying and selling of the merchandise. The device or the system in that office might turn out to be malicious.      

\par
In this paper we propose the concept of ownership authentication transfer in ubiquitous computing devices which takes care of transferring the ownership of the device to the new owner. This is used when an old device is bought by the user. This provides the security for the stolen device and avoids impersonation attacks. It also addresses most of the limitations mentioned above.
\par
Rest of the paper is organized as follows. Section 2 explains the operation of the proposed ownership transfer protocol. Section 3 discusses about the security analysis of the proposed protocol followed by discussion in section 4. Finally section 5 concludes the paper.

\section{Device Ownership Authentication Transfer Protocol}
In this paper we propose a secure and fair protocol for ownership transfer of the ubiquitous computing device. The user who is buying an old device from the other user has to undergo this process in order to successfully acquire the ownership of the device and start using it in the ubiquitous environment.\\

\textbf{Assumption:} The value or the price of the device has been agreed upon between the users before transferring the ownership of the device.\\

\textbf{Requisite:} The users should be in physical proximity and the whole process has to be carried out in the device which is under sale.

\begin{table*}[bpht!]
\centering
	\caption{Notations Used} 
	\label{tab}
	\begin{tabular}{|c|p{2in}|c|p{2in}|}
	\hline 
	\textbf{Symbol}&\textbf{Meaning}&\textbf{Symbol}&\textbf{Meaning} \\ \hline 
			
			$E_{P_{CKS}}$&Encryption Using Public key of CKS&$CKS$&Central Key Server\\ \hline 
			$N_A$&nonce generated by A& $N_B$&Nonce generated by B \\ \hline
			$ID_A$&User ID or User Name of the user A&$ID_B$&User ID or User Name of the user B\\ \hline 
			$P_{CKS}$&Public key of the CKS&$Ack$& Acknowledgment\\ \hline
			$PW_A$&Password of the User A&$Temp ID$& Temporary ID or Pseudo ID\\ \hline 
			$OTC$&Ownership Transfer Confirmation&$OTR$&Ownership Transfer Request\\ \hline
		  
	\end{tabular}
\end{table*}	  

\begin{figure*}[bpht!]
\centering
\includegraphics[width=6.5in,height=4.5in]{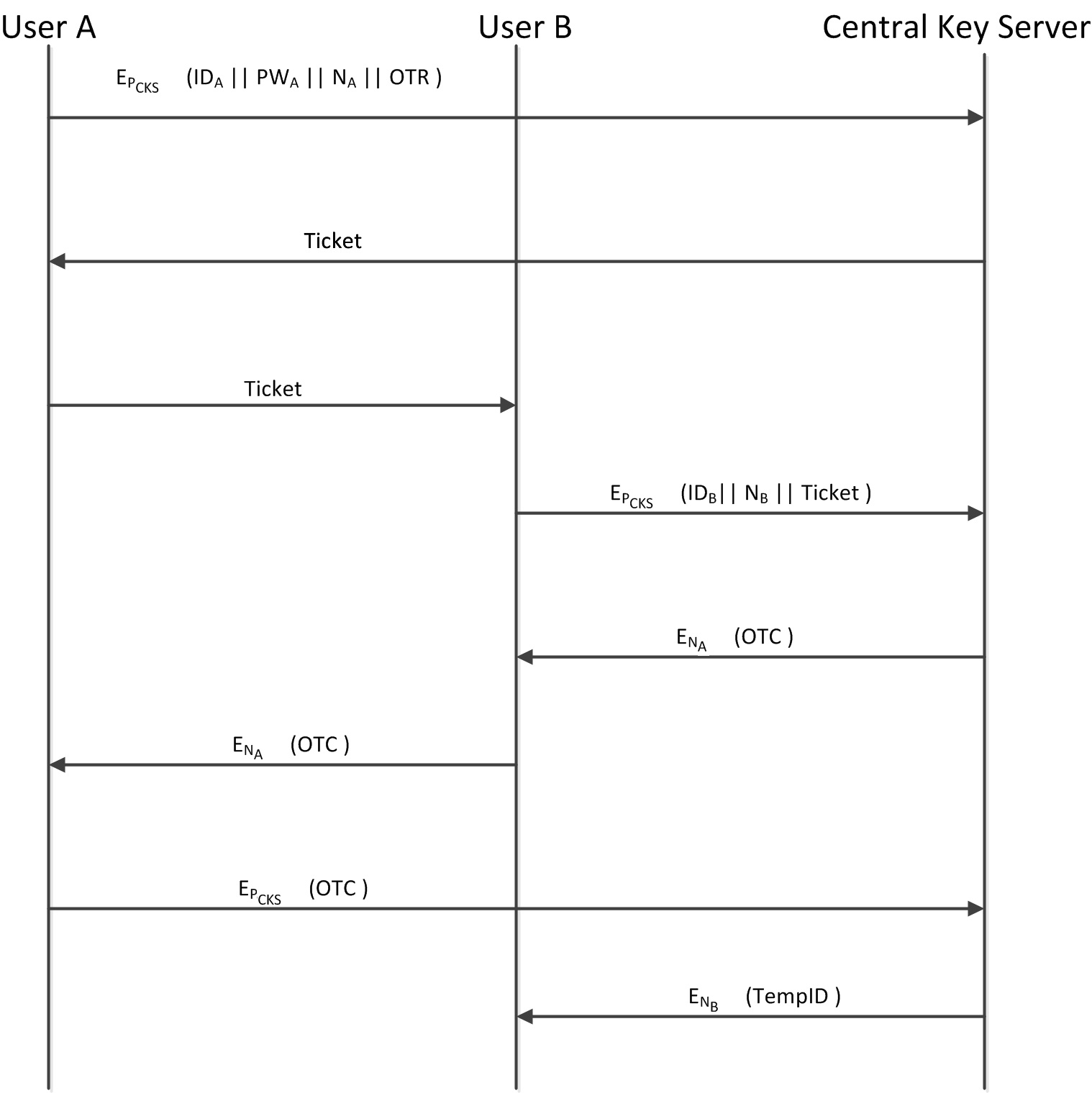}
\caption{Diagram Showing Device Ownership Transfer Process}
\label{fig:1}
\end{figure*}
	
The previously existing user should introduce the new owner of the device to the CKS, in other words user A must transfer the ownership authentication credentials to the new user B. Once the new owner is introduced, the CKS will delete the credentials of the previous owner and then save the credentials of the new owner for the same device. Once the ownership rights has been transfered to the new user, the old user at any cost will not be able to use the device. If in case the whole process of ownership transfer as mentioned below is not completed, either of the users will not be able to use the device. This also takes care of the scenario that if a device is stolen, the thief cannot use the device. The proposed ownership transfer protocol for a given ubiquitous device has been explained below.
			
			\begin{enumerate}
				\item $ U_A\rightarrow CKS \ :\ E_{P_{CKS}} (ID_{A} \| PW_A \| N_A\| OTR)$ \\ 
				The user A (Old User) sends the message to the CKS. The message consists of the user A ID, password of the user A, nonce of the user A and Ownership Transfer Request (OTR). This message is encrypted using the public key of the CKS. OTR consists of the ID of the user selling the device, ID and nonce of the user buying the device. OTR is also encrypted using the public key of the CKS, where $OTR = E_{P_{CKS}} (ID_A \| ID_B \| N_B)$. In this step the user A will introduce user B to the CKS.  
				\item  $CKS \rightarrow U_A \ :\ Ticket$\\ 
				In response to the user A's request for ownership transfer, the CKS sends a ticket to the user A. The Ticket consists of the acknowledgment for ownership transfer to the user B. The ticket is encrypted using the public key of the CKS.
				\item $U_B \rightarrow CKS \ :\ E_{P_{CKS}}(ID_{B} \| Ticket \| N_B)$\\ 
				The user A will now hand over the device to the user B. Now the user B sends his/her credentials to the CKS. The user needs to send user ID, nonce and the ticket got by the user A. The ticket will be in the device itself.   
			 \item $CKS \rightarrow U_B \ :\ E_{N_{A}}(\mbox{Ownership Transfer Confirmation})$\\ 
					Once the CKS receives the credentials of User B, the CKS sends the ownership transfer confirmation to the user B by encrypting it using nonce of the user A. This message consists of the information about the money to be transferred and the account details of the destination account. 
				\item $U_A \rightarrow CKS \ :\  E_{P_{CKS}}(\mbox{Ownership Transfer Confirmation})$\\ 
				The user B will hand over the device to the user A and the user A will decrypt the message, read the acknowledgment and then he/she sends the acknowledgment back to the CKS by encrypting it using the public key of CKS. By sending the acknowledgment back to the CKS, he/she confirms the ownership transfer of the device. Signing a particular message twice is required to strike the fairness in the deal. There may be some chances where either of the users may turn to be malicious. This is done in order to obtain a confirmation from the user who is selling the device. 
				\item $CKS \rightarrow U_B \ :\ E_{N_{B}}(TempID)$\\ 
				On receiving the message, CKS completes the ownership transfer of the device by sending the temp ID  to the user B. The temp ID is encrypted using the nonce of the user B. 
								
			\end{enumerate}
	
	The above explained process of device ownership transfer is summarized in the Fig. \ref{fig:1}.

	
\section{Security Analysis}
The protocol has been modeled for validation using Automated Validation of Internet Security Protocols and Applications (AVISPA) \cite{avispa} \cite{avispapro} \cite{lv}. AVISPA is a tool aimed at developing a push-button, industrial-strength technology to analyze a large-scale Internet security-sensitive protocols and applications. This technology speeds up the development of the next generation of network protocols, improve their security features, and therefore increase the public acceptance of advanced, distributed IT applications based on them. AVISPA is a commonly used verification tool to analyze the cryptographic features of the protocols. AVISPA provides a language known as the High Level Protocol Specification Language (HLPSL) \cite{hlpsl} to describe the security protocols and to specify their intended security properties, as well as a set of tools to formally validate them. It provides a modular and expressive formal language for specifying protocols and their security properties, and integrates different back-ends that implement a variety of state-of-the-art automatic analysis techniques. In order to analyze the proposed technique, the HLPSL specification for Ownership Authentication Transfer is as given below.

\begin{verbatim}

	role usera(
	UA,C: agent,
	Kcks : public_key,
	SND,RCV : channel(dy))
	played_by UA def=
	local
	State : nat,
	Na: text,
	Ida,Pwa,Otr,Otc: message
	init
	State := 0
	transition
	1. State = 0 /\ RCV(start) =|> 
	State':= 2 /\ Na' := new()
	/\ SND({Ida.Pwa.Otr.Na'}_Kcks)
	2.State=2/\RCV({Otc}_Na)
	/\request(UA,C,usera_server_na,Na') =|> 
	State':=8/\SND({Otc}_Kcks)
	end role
	----------------------------------------
	role ck(
	UA,C,UB : agent,
	Kcks : public_key,
	SND,RCV : channel(dy))
	played_by C def=
	local
	State : nat,
	Na,Nb: text,
	Ida,Pwa,Otr,Otc,Tempid,Idb,T: message
	init
	State :=1
	transition
	1.State =1/\ RCV({Ida.Pwa.Otr.Na'}_Kcks)=|> 
	State':=3/\SND(T) 
	2.State=3/\RCV({Idb.T.Nb'}_Kcks)=|> 
	State':=7/\SND({Otc}_Na)
	/\witness(C,UA,usera_server_na,Na)
	3.State=7/\RCV({Otc}_Kcks)=|> 
	State':=9/\SND({Tempid}_Nb)
	/\witness(C,UB,userb_server_nb,Nb)
	end role
	----------------------------------------
	role userb(
	C,UB : agent,
	Kcks: public_key,
	SND,RCV : channel(dy))
	played_by UB def=
	local
	State : nat,
	Nb: text,
	Otc,Tempid,T,Idb: message
	init
	State := 4
	transition
	1. State = 4 /\ RCV(T) =|> 
	State':= 8/\ Nb':=new()
	/\SND({Idb.T.Nb'}_Kcks)
	2.State=8/\RCV({Tempid}_Nb)
	/\request(UB,C,userb_server_nb,Nb')=|>	
	State':=10
	end role
	-----------------------------------
	role session(
	UA,C,UB: agent,
	Kcks : public_key)
	def=
	local SA, SB, RA, RB : channel (dy)
	composition
	usera(UA,C,Kcks,SA,RA) 
	/\ userb(C,UB,Kcks,SB,RB)
	/\ ck(UA,C,UB,Kcks,SB,RB)
	end role
	-------------------------------------
	role environment()
	def=
	const
	userb_server_nb,usera_server_na: protocol_id,
	ks,ki: public_key,
	a,b,c : agent
	intruder_knowledge = {a,b,c,ks,ki}
	composition
	session(a,b,c,ks)
	end role
	------------------------------------------
	goal
	authentication_on userb_server_nb
	authentication_on usera_server_na
	end goal
	------------------------------------------
	environment()
\end{verbatim} 

\begin{figure}[bpht!]
\centering
\includegraphics[width=3in,height=2.5in]{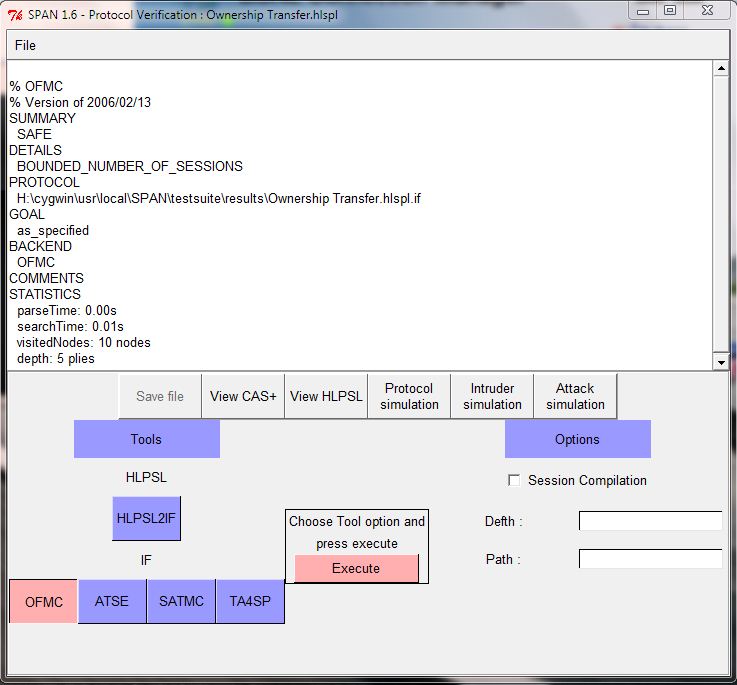}
\caption{Screen Shot Showing the Details of the Result for Ownership Authentication Transfer Phase }
\label{fig:2}
\end{figure}

\begin{figure}[bpht!]
\centering
\includegraphics[width=3in,height=2.5in]{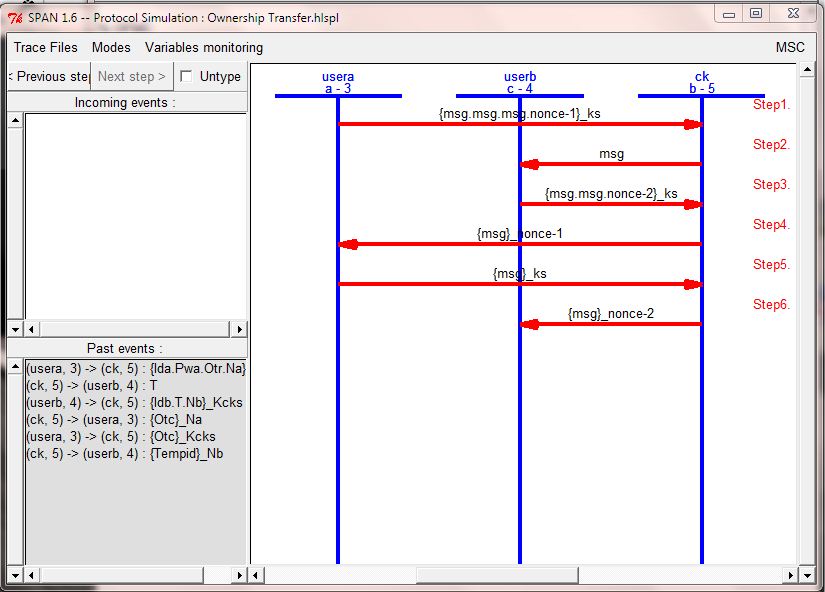}
\caption{Screen Shot Showing the Protocol Simulation for Ownership Authentication Transfer }
\label{fig:3}
\end{figure}

The result of the protocol simulation and verification for the Ownership Authentication Transfer is shown in the Fig.\ref{fig:2} and Fig.\ref{fig:3}

\begin{figure}[bpht!]
\centering
\includegraphics[width=2.85in,height=3in]{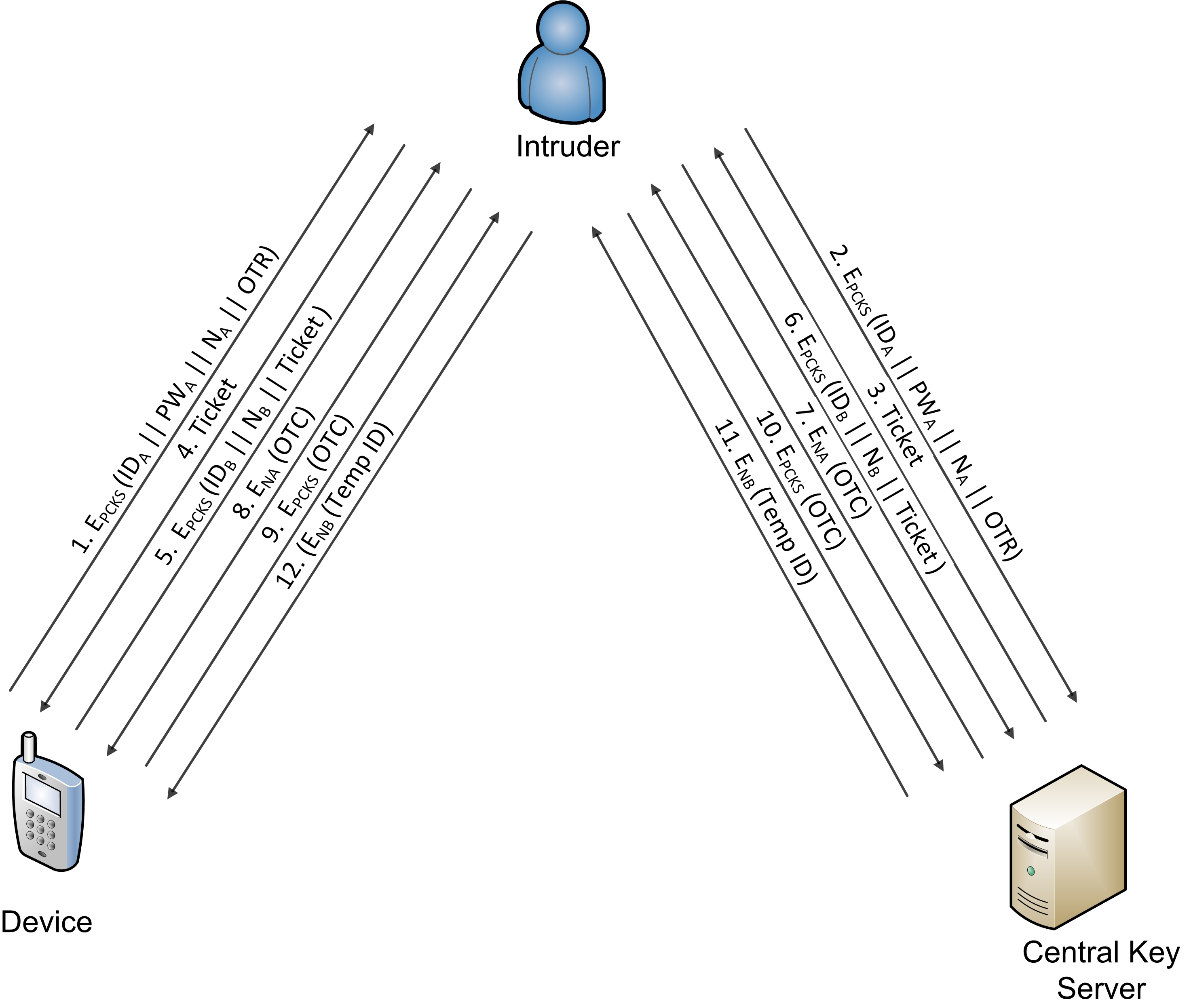}
\caption{Diagram Showing Intruder attack on Ownership Authentication Transfer of the Devices}
\label{fig:4}
\end{figure}

In this phase the old user requests the CKS to start the ownership transfer of the device. Both the new user and the old user will use the same device to interact with the CKS. The users will feed the details into the device that has to be sent to the CKS. The messages that are going out of the device may be intercepted by the intruder. But the intruder will not be able to decrypt the message and know its contents as he/she will not be having the corresponding keys to decrypt the message. Thus again the intruder attack fails and there is no harm caused from the intruder to any of the authenticated entity in the ubiquitous environment. 
The fifth and final HLPSL specification formalizes AOT, in which there are two protocol roles: Device D and Central Key Server C. The intruder I behaves either as D or as C, thus deceiving the other entities in the network. But this kind of attack is not harmful as the intruder will not be able to read the contents of the messages being exchanged.
\begin{enumerate}
	\item $D \rightarrow I ("C") : E_{P_{CKS}} (ID_{A} \| PW_A \| N_A\| OTR)$
	\item $C \rightarrow I ("D") : Ticket$
	\item $D \rightarrow I ("C") : E_{P_{CKS}}(ID_{B} \| Ticket \| N_B)$
	\item $C \rightarrow I ("D") : E_{N_{A}}(\mbox{OTC})$
	\item $D \rightarrow I ("C") : E_{P_{CKS}}(\mbox{OTC})$
	\item $C \rightarrow I ("D") : E_{N_{B}}(TempID)$
\end{enumerate}
The intruder attack is show in the Fig. \ref{fig:4}. In this figure, we have shown only the device under sale instead of the two users for better understanding. 

\section{Discussion}
The Ubicomp device contains sensitive user data and also an easy access to the data which are stored in the CKS, it is important that the user's data is protected. It is not secure for the user to borrow or lend his/her device for any kind of transaction. The above proposed concept provides the transfer of ownership devices securely for the users interested in buying the old devices. If either of the users do not provide accurate information, the transfer will be aborted. At this point, there may be a question arises as to how often the ownership transfer can be done and can there be any temporary ownership transfer. How often the ownership is transferred depends on the user of that device. In developing countries people may tend to buy second hand devices, as the price of the devices will meet their budget. The ownership transfer can be done any number of times based on the interest of the users. On the other hand, the mobiles or the hand held devices are sold and the ownership is transferred permanently. There is nothing like temporary change in the ownership. Even though the there is a need to change the ownership temporarily, it depends on the mutual agreements between the new and the old user. Temporary ownership transfer is purely at the users risk and it is not a part of this protocol. This concept also takes care of the scenario where the device has been lost or stolen. No person will be able to use the device other than its owner. Thus it avoids impersonation attack.
\par
The protocol is verified using Automated Validation of Internet Security Protocols and Applications (AVISPA). AVISPA provides a language known as the High Level Protocol Specification Language (HLPSL) to describe the security protocols and to specify their intended security properties, as well as a set of tools to formally validate them. Experiments that were conducted on the vast library of Internet security
protocols have indicated that the AVISPA Tool is a leading edge tool for Internet security protocol analysis. There are no other tool that exhibit the same level of scope and robustness providing the excellent performance and scalability to the best of our knowledge. The proposed protocol was modeled and tested using AVISPA tool and was found to be safe.

\section{Conclusion}
By incorporating the concept of ownership authentication transfer in the ubiquitous environment, we provide more security with respect to the owner's sensitive data and also to his device. The device can only be accessed by its owner. If the device is stolen or lost the device cannot be accessed by the other users unless the process of ownership transfer is completely done. Thus this adds up the security in the ubiquitous environment. This also avoids the impersonation attack. 

\vspace{1cm}
\bibliographystyle{IEEEtran}
\bibliography{ref}
\end {document}